\documentclass[superscriptaddress,twocolumn,prx,preprintnumbers,amsmath,amssymb,longbibliography]{revtex4}
\usepackage{graphicx,afterpage}
\usepackage{xcolor}
\usepackage{ulem}
\usepackage[colorlinks=true,plainpages=false,linkcolor=blue,urlcolor=blue,citecolor=blue,pdfpagemode=UseNone,pdfstartview=FitBH]{hyperref}
\usepackage[utf8]{inputenc}

\DeclareMathAlphabet{\altmathcal}{OMS}{cmsy}{m}{n}

\begin{document}

\title{Spin-stripe order tied to the pseudogap phase in La$_{1.8-{\rm x}}$Eu$_{0.2}$Sr$_{\rm x}$CuO$_4$}

\author{A.~Missiaen}
\affiliation{CNRS, LNCMI, Univ. Grenoble Alpes, INSA-T, UPS, EMFL, Grenoble, France}
\author{H.~Mayaffre}
\affiliation{CNRS, LNCMI, Univ. Grenoble Alpes, INSA-T, UPS, EMFL, Grenoble, France}
\author{S.~Kr\"amer}
\affiliation{CNRS, LNCMI, Univ. Grenoble Alpes, INSA-T, UPS, EMFL, Grenoble, France}
\author{D.~Zhao}
\affiliation{Hefei National Laboratory for Physical Sciences at the Microscale, University of Science and
Technology of China, Hefei, Anhui 230026, China}

\author{Y.B.~Zhou}
\affiliation{Hefei National Laboratory for Physical Sciences at the Microscale, University of Science and
Technology of China, Hefei, Anhui 230026, China}

\author{T.~Wu}
\affiliation{Hefei National Laboratory for Physical Sciences at the Microscale, University of Science and
Technology of China, Hefei, Anhui 230026, China}
\affiliation{CAS Key Laboratory of Strongly-coupled Quantum Matter Physics, Department of Physics,
University of Science and Technology of China, Hefei, Anhui 230026, China}
\affiliation{Collaborative Innovation Center of Advanced Microstructures, Nanjing University, Nanjing 210093, China}

\author{X.H.~Chen}
\affiliation{Hefei National Laboratory for Physical Sciences at the Microscale, University of Science and
Technology of China, Hefei, Anhui 230026, China}
\affiliation{CAS Key Laboratory of Strongly-coupled Quantum Matter Physics, Department of Physics,
University of Science and Technology of China, Hefei, Anhui 230026, China}
\affiliation{Collaborative Innovation Center of Advanced Microstructures, Nanjing University, Nanjing 210093, China}

\author{S. Pyon}
\affiliation{Department of Advanced Materials Science, University of Tokyo, Kashiwa 277-8561, Japan}
\affiliation{Department of Applied Physics, University of Tokyo, Tokyo 113-8656, Japan}

\author{T. Takayama}
\affiliation{Department of Advanced Materials Science, University of Tokyo, Kashiwa 277-8561, Japan}
\affiliation{Max Planck Institute for Solid State Research, Stuttgart 70569, Germany}

\author{H. Takagi}
\affiliation{Department of Advanced Materials Science, University of Tokyo, Kashiwa 277-8561, Japan}
\affiliation{Max Planck Institute for Solid State Research, Stuttgart 70569, Germany}
\affiliation{Department of Physics, University of Tokyo, Tokyo 113-0033, Japan}
\affiliation{Institute for Functional Matter and Quantum Technologies, University of Stuttgart, Stuttgart 70049, Germany}

\author{D. LeBoeuf}
\affiliation{CNRS, LNCMI, Univ. Grenoble Alpes, INSA-T, UPS, EMFL, Grenoble, France}

\author{M.-H. Julien}
\email{marc-henri.julien@lncmi.cnrs.fr}
\affiliation{CNRS, LNCMI, Univ. Grenoble Alpes, INSA-T, UPS, EMFL, Grenoble, France}

\date{\today}

\begin{abstract}
Although spin and charge stripes in high-$T_c$ cuprates have been extensively studied, the exact range of carrier concentration over which they form a static order remains uncertain, complicating efforts to understand their significance. \textcolor{black}{The problem is challenging due to the combined effects of quenched disorder and competition with superconductivity --~both significant in cuprates~-- which add to the inherent difficulty of determining phase boundaries}. In La$_{2-{\rm x}}$Sr$_{\rm x}$CuO$_4$ (LSCO) and in zero external magnetic field, static spin stripes are confined to a doping range well below $p^*$, the pseudogap boundary at zero temperature. However, when high fields suppress the competing effect of superconductivity, spin stripe order is found to extend up to $p^*$. Here, we investigated La$_{1.8-{\rm x}}$Eu$_{0.2}$Sr$_{\rm x}$CuO$_4$ (Eu-LSCO) using $^{139}$La nuclear magnetic resonance and observe field-dependent spin fluctuations suggesting a similar competition between superconductivity and spin order as in LSCO. Nevertheless, we find that static spin stripes are present practically up to $p^*$ irrespective of field strength: the stronger stripe order in Eu-LSCO prevents superconductivity from enforcing a non-magnetic ground state, except very close to $p^*$. Thus, spin-stripe order is consistently bounded by $p^*$ in both LSCO and Eu-LSCO, despite their differing balances between stripe order and superconductivity. This indicates that the canonical stripe order, where spins and charges are intertwined in a static pattern, is fundamentally tied to the pseudogap phase, \textcolor{black}{though the exact nature of this connection has yet to be elucidated}. Any stripe order beyond the pseudogap endpoint must then be of a different nature: either spin and charge orders remain intertwined, but both fluctuating, or only spin order fluctuates while charge order remains static. The presence of spin-stripe order up to $p^*$, the pervasive, slow, and field-dependent spin-stripe fluctuations, as well as the electronic inhomogeneity documented in this work, must all be carefully considered in discussions of Fermi surface transformations, putative quantum criticality, and strange metal behavior.

\end{abstract}
\maketitle

\section{Introduction}

The phase diagram of high $T_c$ cuprates as a function of temperature $T$ and hole doping $p$ features various electronic phases that remain insufficiently understood~\cite{Keimer2015,Proust2019}. Determining the boundaries of these phases is essential to unravel possible connections between them as well as with the superconducting phase. 

The boundary of the pseudogap phase at $T=0$ is considered to be well-defined in a number of cuprate families, terminating at a doping $p^*$. The exact value of $p^*$ is somewhat compound dependent, but it essentially lies around 0.2 doping, in-between $p\simeq 0.16$, the optimal doping for superconductivity, and $p\simeq 0.3$, the end of the superconducting dome. The boundaries of spin or charge ordered states, on the other hand, are in general far less certain and this uncertainty hinders the understanding of the pseudogap state, whose nature continues to be actively debated. In particular, there can hardly be a thorough discussion of putative quantum criticality~\cite{Ramshaw2015,Michon2019,Arpaia2023} until the boundaries of the ordered phases at work in the cuprate phase diagram are precisely known.

Consider the cuprates La$_{1.8-{\rm x}}$Eu$_{0.2}$Sr$_{\rm x}$CuO$_4$ (Eu-LSCO) and La$_{1.6-{\rm x}}$Nd$_{0.4}$Sr$_{\rm x}$CuO$_4$ (Nd-LSCO), two members of the broader family of lanthanum-based cuprates (La214). Both are characterized by nearly identical low temperature tetragonal (LTT) phase transitions, prominent spin and charge stripe orders around ${\rm x} = p = 0.12$ doping as well as concomitant weakening of three-dimensional superconductivity due to the competition between superconducting and stripe orders (Fig.~1 and ref.~\cite{Huecker2012}). The two compounds are considered to have essentially identical magnetic and electronic phase diagrams, including a pseudogap boundary occurring at a doping $p^*\simeq 0.235 \pm 0.005$ hole/Cu according to ref.~\cite{Cyr2018}. Recently, it has been suggested that $p^*$ in Nd/Eu-LSCO represents a quantum critical point~\cite{Michon2019} that governs important electronic properties of this compound, including a sharp change in the Fermi surface topology and the carrier density across $p^*$~\cite{Fang2022,Collignon2017,Gourgout2022}. To interpret these observations, it is important to determine where the spin and charge orders terminate in the phase diagram. 

The cuprate stripes have been extensively studied over the last three decades. It may thus appear surprising that their $T=0$ boundaries as a function of hole doping, in particular the upper boundary, are not accurately known. In reality, this question is more involved than it may seem: besides the obvious fact that electronic modulations become inherently weak upon approaching the boundaries, there are various difficulties associated with the complex physics of the cuprates. First, because quenched disorder from dopants has severe effects on charge degrees of freedom, spin and charge orders show a substantial degree of spatial heterogeneity together with glass-like dynamics involving wide distributions of fluctuation frequencies~\cite{Kivelson1998,Julien1999,Kohsaka2007,Tranquada2020,Vinograd2022}. In this situation, distinguishing truly static order from fluctuating order, if even feasible, requires subtle consideration of experimental parameters such as the measurement timescale or the energy resolution. Second the stripe modulations can be weakened by their competition with superconductivity at low temperatures. More critically, superconductivity may even entirely suppress stripe order, which then makes its boundary magnetic-field dependent, as observed in La$_{2-{\rm x}}$Sr$_{\rm x}$CuO$_4$ (LSCO), a sibling compound that lacks the LTT transition but nonetheless also shows stripe order~\cite{Frachet2020,Vinograd2022}. This example, as well as the short-ranged to long-ranged CDW transition in YBa$_2$Cu$_3$O$_{\rm y}$~\cite{Julien2015}, highlight a general issue: as essential as it is for revealing underlying boundaries, the field is not a simple knob that just removes superconductivity. Electronic properties are field-dependent, in some cases even after superconductivity is removed~\cite{Frachet2021,Vinograd2022}, and the high-field ground state may differ from the zero-field one. This obviously complicates the comparison between experiments performed with and without field. 

With these pitfalls, it is perhaps not surprising that the boundaries of stripe order in La214 have remained unsettled. In Nd-LSCO with ${\rm x}=0.20$, the lack of signature  of spin order in a muon-spin rotation ($\mu$SR) experiment~\cite{Nachumi1998} has suggested that stripes have disappeared before $p^*$, thus supporting the view~\cite{Michon2019} that the electronic changes occurring at $p^*$ are unrelated to the presence or absence of stripes. However, other studies have found differently: signatures of magnetic order have been reported in $\mu$SR and nuclear magnetic resonance (NMR) studies of Eu-LSCO ${\rm x}\simeq0.2$~\cite{Klauss2000,Suh2000,Hunt2001,Grafe2010} as well as in neutron scattering studies of Nd-LSCO ${\rm x}\simeq0.2$~\cite{Tranquada1997} and even ${\rm x}\simeq0.24$~\cite{Ma2021}. 

The situation is equally confusing concerning the charge sector: a study of Nd-LSCO found no CDW order for ${\rm x} \geq 0.18$~\cite{Gupta2021} while another study of Eu-LSCO found a clear CDW peak at ${\rm x}=0.20$~\cite{Lee2022}.  In LSCO, X-ray scattering studies have led to similarly conflicting conclusions regarding the end doping of CDW order, either before, at or above $p^*$~\cite{Wen2019,Miao2021,vonArx2023,Li2023}. At the same time, thermopower measurements in LSCO~\cite{Badoux2016} and Nd-LSCO~\cite{Collignon2021} have been interpreted as evidence that CDW order ends well before $p^*$, even in high field.

   \begin{figure}[t!]
 \includegraphics[width=8.5cm]{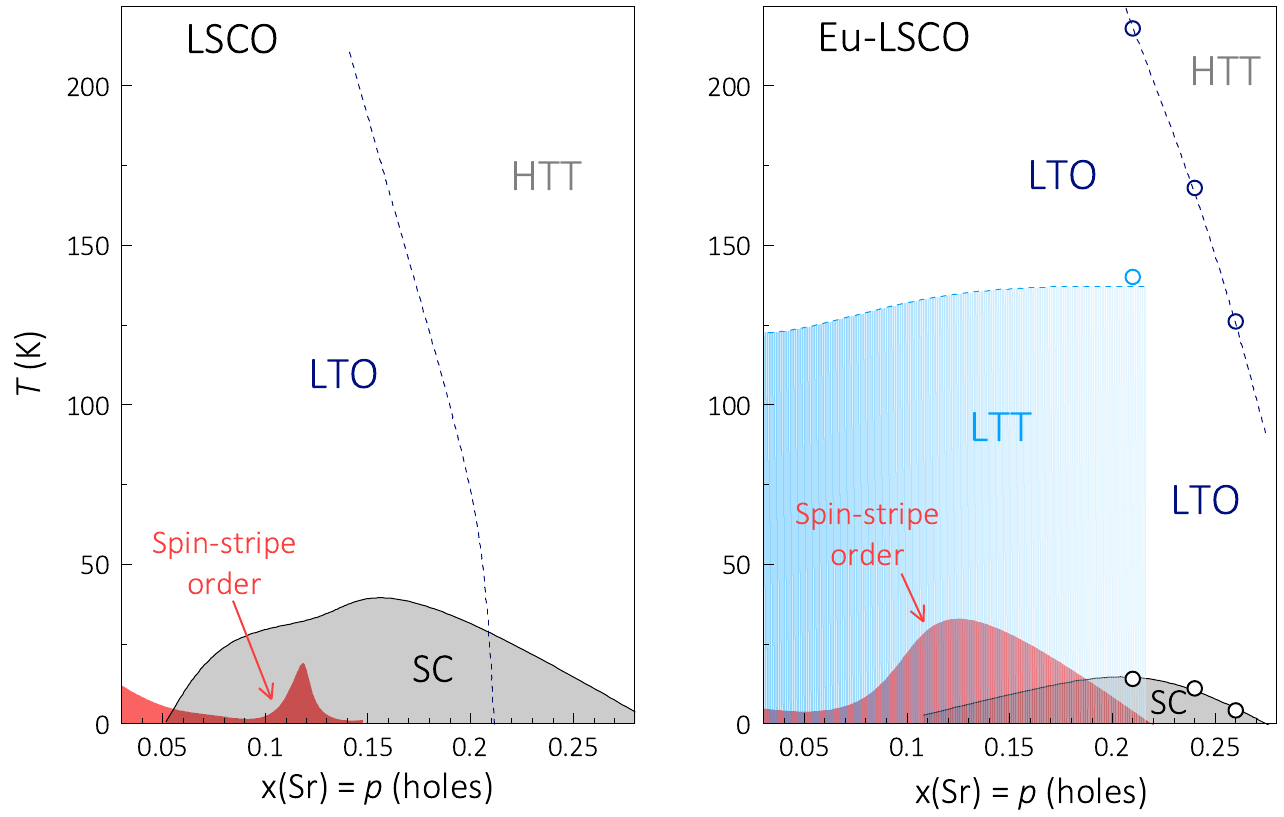}
  \caption{\label{phasediag} Zero-field phase diagrams of La$_{2-{\rm x}}$Sr$_{\rm x}$CuO$_4$ (LSCO) and La$_{1.8-{\rm x}}$Eu$_{0.2}$Sr$_{\rm x}$CuO$_4$ (Eu-LSCO). HTT (high-temperature tetragonal) and LTO (low-temperature orthorhombic) phases are common to the two compounds (though with different boundaries) whereas the LTT (low-temperature tetragonal) phase \textcolor{black}{-- colored in light blue --} is specific to Eu-LSCO.  Grey and red areas depict the superconducting (SC) and spin-ordered phases in zero field, respectively. The phase boundaries are taken from refs.~\cite{Klauss2000,Julien2003,Frachet2020}. Open circles in (b) correspond to measurements on the three samples of this study (see Fig.~\ref{tuning} for $T_c$ measurements and Appendix for the determination of the structural phase transitions from $1/T_1$ measurements). }
\end{figure} 

In this article, we present NMR measurements on Eu-LSCO single crystals, aimed at detecting spin-stripe order at doping levels below, at, and above $p^*$, under conditions where superconductivity is either present or suppressed by strong magnetic fields. Eu-LSCO offers an advantage over Nd-LSCO, as Eu$^{2+}$ has a nonmagnetic ground state. In contrast, in Nd-LSCO, fluctuations and ordering of the Nd$^{3+}$ moments can potentially obscure the intrinsic response from the CuO$_2$ planes~\cite{Tranquada1997,Hunt2001}.

\section{NMR and spin freezing}

Spin-stripe order is detected in the $^{139}$La nuclear spin-lattice relaxation rate $1/T_1$ that probes the spectral density of spin fluctuations at very low energy, specifically at the NMR frequency $\omega_{n} \sim$~MHz~$\sim \mu$eV (see~\cite{Vinograd2022} and references therein).  As fluctuations slow down with cooling, the spectral weight shifts to progressively lower energy, which results in a peak of $1/T_1$ at the temperature where the spectral density at $\omega_{n}$ is maximized. This temperature defines the freezing temperature on the NMR timescale, $T_{\rm g}^{\rm NMR}$. In La214, this freezing process is characterized by spatial inhomogeneity, leading to a substantial distribution of $T_1$ values whose exact form has been the subject of recent discussion~\cite{Arsenault2020,Singer2020,Vinograd2022}. This heterogeneity, along with the successful description of the data in terms of gradual slowing down across $\omega_{n}$ and $T_{\rm g}^{\rm NMR}$, have led to qualify the freezing as "glassy".

Quantitatively speaking, the broad peak of $T_1^{-1}~vs.~T$ in striped cuprates is captured by a simple model assuming that:

(1) The dynamics can be described by a correlation time $\tau_c$ that becomes exponentially long upon cooling:
\begin{equation}
\tau_{c}(T) = \tau_{\infty}\,e^{E_{0}/k_B T} \,.
\label{eq:tau}
\end{equation}

(2) The spin dynamical structure factor 
\begin{equation}
\altmathcal{S}_{\perp}(\omega)=   \frac{1}{2\pi}\int_{-\infty}^{+\infty} \langle S_+(t)S_-(0) \rangle e^{i\omega t} \, dt \, ,
\label{eq:Sperp}
\end{equation}
(or, more exactly, its integral over momenta (${\bf q}$) weighted by a ${\bf q}$-dependent form factor) is a Lorentzian function of width $\tau_c^{-1}$, centered at $\omega=0$:
\begin{equation}
\altmathcal{S}_{\perp}(\omega)=   S_\perp^2 \, \tau_c/(1+(\omega \tau_{c})^2) \, , 
\label{eq:tau2}
\end{equation}
 where $S_\perp^2 = \langle S_+(0)S_-(0)\rangle=\int \altmathcal{S}_\perp(\omega)d\omega$ is the fluctuating moment squared. $T^{-1}_{1}$ is proportional to the low-frequency  ($\omega=\omega_{n}$) limit of $\altmathcal{S}_{\perp}(\omega)$, which gives:
\begin{equation}
T^{-1}_{1} \propto   S_\perp^2 \frac{\tau_{c}}{1+(\omega_{n} \tau_{c})^2} .
\label{eq:BPP}
\end{equation}

Under these assumptions, Eq.~\ref{eq:BPP} peaks at $T=T_g^{\rm NMR}$, the temperature at which $\tau_c^{-1}=\omega_n$, thereby defining the freezing temperature on the NMR timescale.

In NMR, $\omega_{n}\propto B$ to first order (provided the Zeeman interaction is much stronger than the quadrupole interaction). This has two implications: first, it is difficult to decouple the field dependence from the probe-frequency dependence (contrary to sound-velocity measurements, for example~\cite{Frachet2021}). Second, if $\altmathcal{S}_\perp(\omega)$ (Eq.~\ref{eq:Sperp}) does not depend on $B$, then $1/T_1 \propto B^{-1}$ at the peak temperature ($\omega_{n} \, \tau_c=1$ in Eq.~\ref{eq:BPP}). In La214, the competition between superconductivity and stripe order breaks this proportionality and $1/T_1$ strongly increases with $B$~\cite{Frachet2020,Vinograd2022}.

   \begin{figure*}[t!]
 \includegraphics[width=15cm]{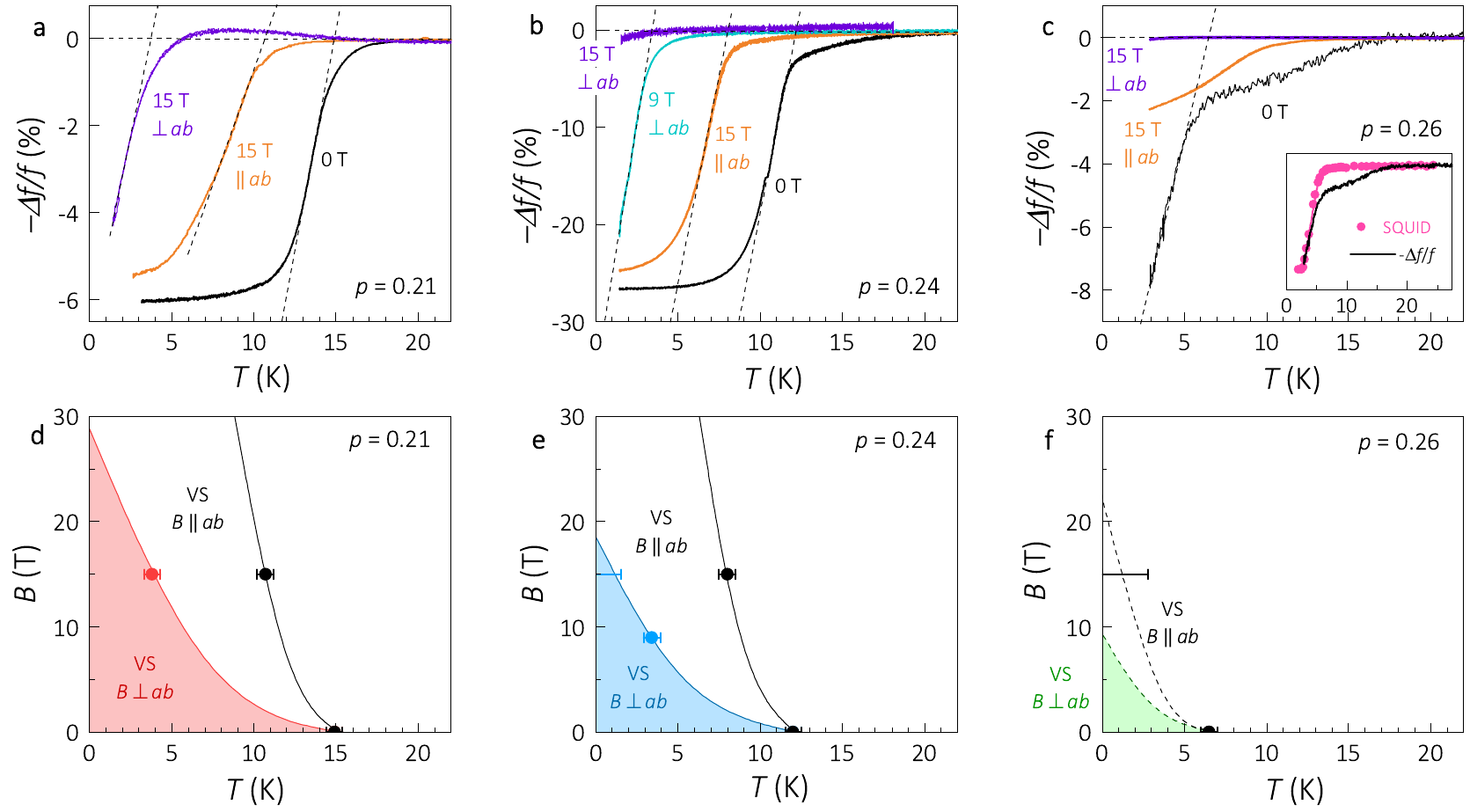}
  \caption{\label{tuning} Superconductivity vs. magnetic field. Top: change $\Delta f$ in resonance frequency ($f$) of the NMR tank circuit as a function of temperature, for different fields, for $p=0.21$ (a), $p=0.24$ (b) and $p=0.26$ (c) samples. Transition temperatures are defined by the intercept between $\Delta f/f$= 0 and the dashed lines describing the linear part of the transition. \textcolor{black}{In zero field, the transition corresponds to the superconducting transition $T_c$ whereas under field, it corresponds to the vortex-melting transition. VS in (d - f) thus stands for "vortex solid". For $p=0.26$ (c), there is a first step below 16~K at 0~T, followed by a sharper change below 6 K. This first step (also seen below 10~K for $B\parallel ab$) is inconsistent with the $T_c$ value expected at $p=0.26$. Inhomogeneity in the bulk cannot explain this step as no evidence of two phases is found in our NMR spectra that are similar to those in LSCO. The effect is attributed to a surface effect as it is not seen in the bulk magnetization (inset to panel c) measured on the same sample with a SQUID. The change in inductance measured with the NMR tank circuit is produced by the change in surface conductivity of the sample~\cite{Zhou2017}, and this measurement is all the more sensitive to the surface that it is performed at MHz frequencies with an NMR coil adjusted to the crystal dimensions. Since 16~K is close to the $T_c$ of the $p=0.21$ sample, the first step in the transition of the $p=0.26$ is ascribed to lower doping at the surface. A similar effect is actually present for $p=0.24$ (panel b). Even slight oxygen absorption at the crystals' edges in ultra-high quality YBa$_2$Cu$_3$O$_{\rm y}$ produces similar steps for this type of measurement}. Bottom: transition temperatures for each field orientation. The lines reproduce the shape of the melting line observed in La$_{1.852}$Sr$_{0.148}$CuO$_4$~\cite{Frachet2020,Vinograd2022}. For $p=0.26$, phase boundaries (dashes) are approximate given the absence of direct evidence of bulk superconductivity at 15~T (c). }
\end{figure*} 

\section{Results}

\subsection{Structural and superconducting properties}

Fig.~\ref{phasediag} shows the superconducting, crystallographic and spin-stripe phases of LSCO and Eu-LSCO in zero field, together with the values of the superconducting and structural transition temperatures for our three Eu-LSCO single crystals: ${\rm x}=0.21$, ${\rm x}=0.24$ and ${\rm x}=0.26$ (see Appendix for more details). Notice that throughout this article, we shall refer to the hole content $p$ of the samples, which is assumed to be equal to the Sr concentration ${\rm x}$. 

The temperatures of the structural transitions are obtained from $1/T_1$ measurements (see Appendix). While all three samples show a transition from the high temperature tetragonal (HTT) phase to the low temperature orthorhombic (LTO), only the $p=0.21$ sample shows an additional LTT phase at low temperature. 

Measurements of the superconducting transition temperature $T_c$ are shown in Fig.~\ref{tuning} (\textcolor{black}{notice that, in a finite field, the measured transition corresponds to the vortex-melting transition). The two-step transition (Fig.~\ref{tuning}c) is observed in all our samples with $p=0.26$, as well as for $p=0.24$ (Fig.~\ref{tuning}b). However, it is absent in our measurements of the magnetization using a quantum interferometer device (SQUID)  (inset to Fig.~\ref{tuning}c). The first step is thus attributed to very small, nearly optimally-doped regions ($p\simeq 0.16$) at the surface (see detailed explanations in the caption to Fig.~\ref{tuning}). This is presumably different from the two-step transition in the magnetization of strongly overdoped LSCO, interpreted as evidence of granular superconductivity arising from electronic inhomogeneity in the bulk~\cite{Li2022,Tranquada2024}.}

\subsection{Considerations on $p^*$ and doping inhomogeneity}

\textcolor{black}{Our $p=0.21$ and 0.24 crystals were cut in the same rods as crystals extensively studied using various techniques, including electrical resistivity~\cite{Michon2019}, specific heat~\cite{Michon2019}, Seebeck and Nernst effects~\cite{Laliberte2015,Cyr2018}, thermal Hall conductivity (ref.~\cite{Chen2024} and references therein), angle-resolved photoemission spectroscopy (ARPES, ref.~\cite{Kuspert2022} and references therein), and X-ray diffraction~\cite{vonArx2023}. Our measured values of the superconducting or structural transitions are in agreement with these studies, as well as with earlier work~\cite{Klauss2000,Huecker2012}. Additionally, all these studies consistently show that the properties of Eu-LSCO are identical to those of Nd-LSCO at the same nominal doping level, whether $p=0.21$ or 0.24. Therefore, while our NMR measurements do not provide direct evidence for the presence or absence of the pseudogap, the overall consistency of results indicates that we can confidently rely on the literature.}

\textcolor{black}{According to the literature, $p=0.21$ lies within the pseudogap phase (with direct evidence from ARPES~\cite{Kuspert2022}) and $p^*=0.235 \pm 0.005$~\cite{Cyr2018} or $0.23\pm0.01$~\cite{Fang2022,Collignon2017,Michon2019} for both Nd-LSCO and Eu-LSCO. This places $p=0.26$ clearly above $p^*$. However, we would not consider $p=0.24$ to be above $p^*$}. This is because even the best La214 crystals ineluctably exhibit a very substantial distribution of local hole concentration: according to Singer {\it et al.}~\cite{Singer2002,Singer2005}, $\Delta p \simeq \pm0.05$ hole near optimal doping. \textcolor{black}{This electronic inhomogeneity has nothing to do with sample quality (and is thus presumably identical in different samples of the same doping) but is intrinsic}, rooted in the electrostatic potential induced by the Sr dopants and/or in an electronic tendency towards phase separation. The smooth evolution of the average properties as a function of doping as well as the very short length scale over which inhomogeneity likely occurs (potentially resembling a kind of electronic micro-emulsion~\cite{Jamei2005,Schmalian2005}) make a direct detection of electronic inhomogeneity challenging.

Inhomogeneity does not concern the sole $p=0.24$ doping but it becomes particularly acute when discussing a sample that lies at, or near, the boundary between two phases: a $p=0.24$ sample likely contains a significant fraction of sites that are effectively below $p^*$, and an equally significant fraction above. This is not contradictory with the same sample having transport properties typical of $p>p^*$ since transport occurs via the most conducting path, {\it i.e.} the highest-doping regions of the samples. On the other hand, NMR is a local probe sensitive to inhomogeneity: the $T_1$ values in the present work represent some average of an unknown distribution of $T_1$ values. Furthermore, the $T_1$ values are probably more representative of the (low doping) regions where magnetism is stronger. In this context, \textcolor{black}{regarding our $p=0.24$ sample as lying above $p^*$ seems inappropriate for the analysis of NMR results.}

   \begin{figure*}[t!]
 \includegraphics[width=17cm]{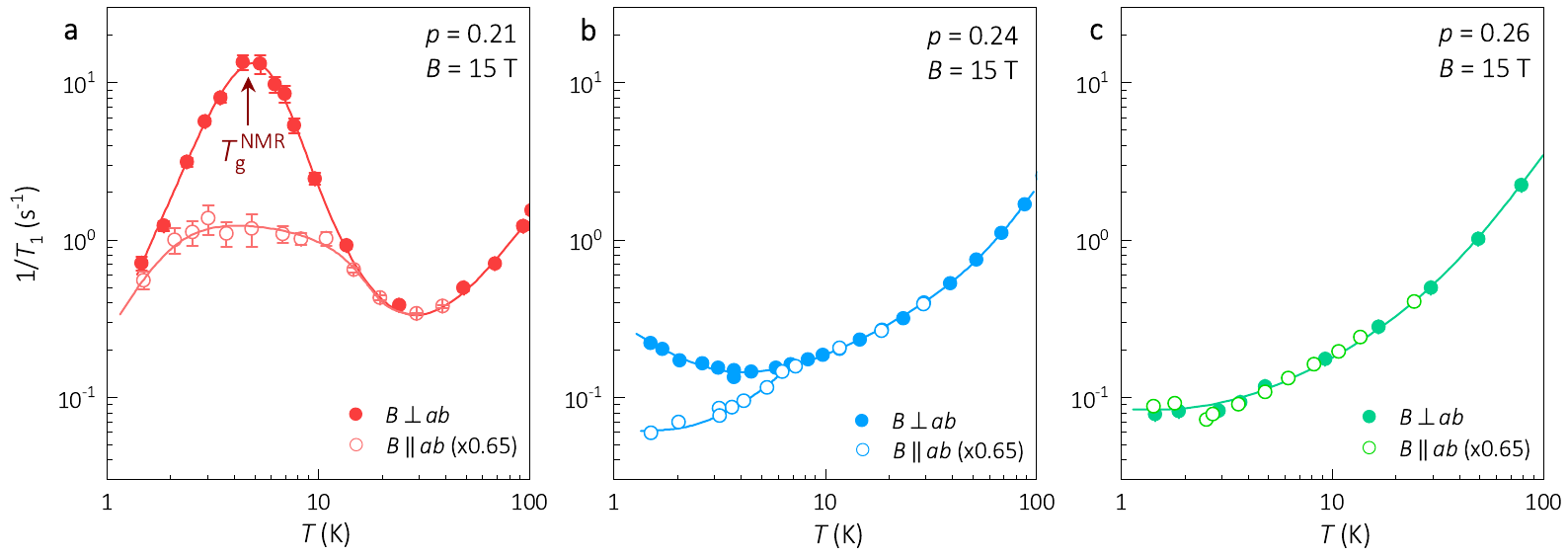}
  \caption{\label{T1angle} Effect of field orientation on spin dynamics. $1/T_1$ vs. temperature in $B=15$~T for $p=0.21$ (a), $p=0.24$ (b) and $p=0.26$ (c), in two perpendicular field orientations. The $T$ and $p$ independent anisotropy factor of 0.65 arises from the anisotropy of the hyperfine coupling. The lines are guides to the eye. }
\end{figure*} 

\subsection{Effect of field orientation on spin order}

Given that a magnetic field $B$ weakens superconductivity much more when applied perpendicular, rather than parallel, to the CuO$_2$ planes, we first test the competition between spin freezing and superconductivity by studying the effect of field orientation at 15  T (Fig.~\ref{T1angle}).

Below $p^*$ ({\it i.e.} for $p=0.21$), the observation of peaks in $1/T_1$ vs. $T$ shows that spin freezing occurs for both orientations (Fig.~\ref{T1angle}a). This is consistent with the above-mentioned evidence of magnetic order already in zero field at this doping level~\cite{Klauss2000,Hunt2001} or in relatively low fields~\cite{Suh2000,Grafe2010}. 

A novel finding here is that the values of $1/T_1$ differ by an order of magnitude between the two field orientations, indicating that spin-stripe order is weaker \textcolor{black}{({\it i.e.}, $S_\perp^2$ in Eq.~\ref{eq:BPP} is smaller)} when superconductivity is stronger. This shows that spin-stripe order coexists and competes with superconductivity at this doping, as it does near $p \simeq 0.12$~\cite{Hunt2001,Mitrovic2008,Arsenault2018}.

 \textcolor{black}{In principle, one would expect the peak temperature $T_{\rm peak}$ to be lower for $B \parallel ab$ than for $B \perp ab$. Indeed, since $T_{\rm peak} = -\frac{E_{0}}{k_B}\ln(\omega_n \tau_{\infty})$ and $\omega_n$ is identical for both field orientations, any difference must arise from $E_0$. Given that $E_0$ is known to anti-correlate with the strength of superconductivity~\cite{Vinograd2022}, it should be much smaller for $B \parallel ab$. Although our $B \parallel ab$ data do not provide clear evidence for a lower $T_{\rm peak}$, the peak is, in fact, ill-defined due to its broad, flat-topped shape. Fitting this data (Fig.~\ref{fits}a) is only feasible if one assumes a considerable distribution of parameters, in which case the average peak position is no longer solely determined by $E_0$. It is likely that when superconductivity is strong, severe spatial inhomogeneity and/or significant deviation from Eq.~\ref{eq:tau} render the fitting model irrelevant. A similar situation was observed in La$_{1.852}$Sr$_{0.148}$CuO$_4$ for low-magnitude out-of-plane fields~\cite{Vinograd2022}.} 
 
At or very near $p^*$ ($p=0.24$), $1/T_1$ shows an upturn upon cooling for $B\perp ab$, signifying that the spectral weight of spin fluctuations at very low energy ($\sim\mu$eV) is increasing at low $T$. Applying a parallel field, however, completely changes the situation: instead of showing a peak, $1/T_1$ drops markedly, as expected for a gapped superconductor (Fig.~\ref{T1angle}b). The absence of increase in $1/T_1$ implies the absence of spin freezing for $B\parallel ab$ and, by extension, in zero field. In this situation, superconductivity is thus strong enough to prevent spin-stripe order. Similarly-striking effect of the field orientation on spin freezing has previously been observed in LSCO~\cite{Frachet2020} and in YBCO~\cite{Wu2013PRB} -- to a somewhat lesser extent for the latter as weak order is already present in zero field. 

Above $p^*$ ($p=0.26$), $1/T_1$ is identical for both fields orientations, to within a $T$ and $p$ independent anisotropy factor of 0.65 (Fig.~\ref{T1angle}c). The absence of \textcolor{black}{gap-like behavior and of} orientation dependence are consistent with the possible absence of bulk superconductivity down to at least 3~K at 15~T (Fig.~\ref{tuning}c). In simple metals, $1/T_1 \propto T$ whereas here $1/T_1$ saturates to a finite value as $T\rightarrow 0$ (Fig.~\ref{T1angle}c). We interpret this saturation as well as the the distribution of $T_1$ values (stretching exponent $\beta\simeq 0.6$ at low $T$,  Fig.~\ref{stretching}) as evidence of nuclear relaxation driven by the temporal fluctuations of inhomogeneous electronic moments. These could originate from clusters of staggered moments around non-magnetic defects~\cite{Julien2000} or from small patches with remaining stripe correlations (see discussion below). 

\begin{figure*}[t!]
\vspace{3mm}
 \includegraphics[width=17cm]{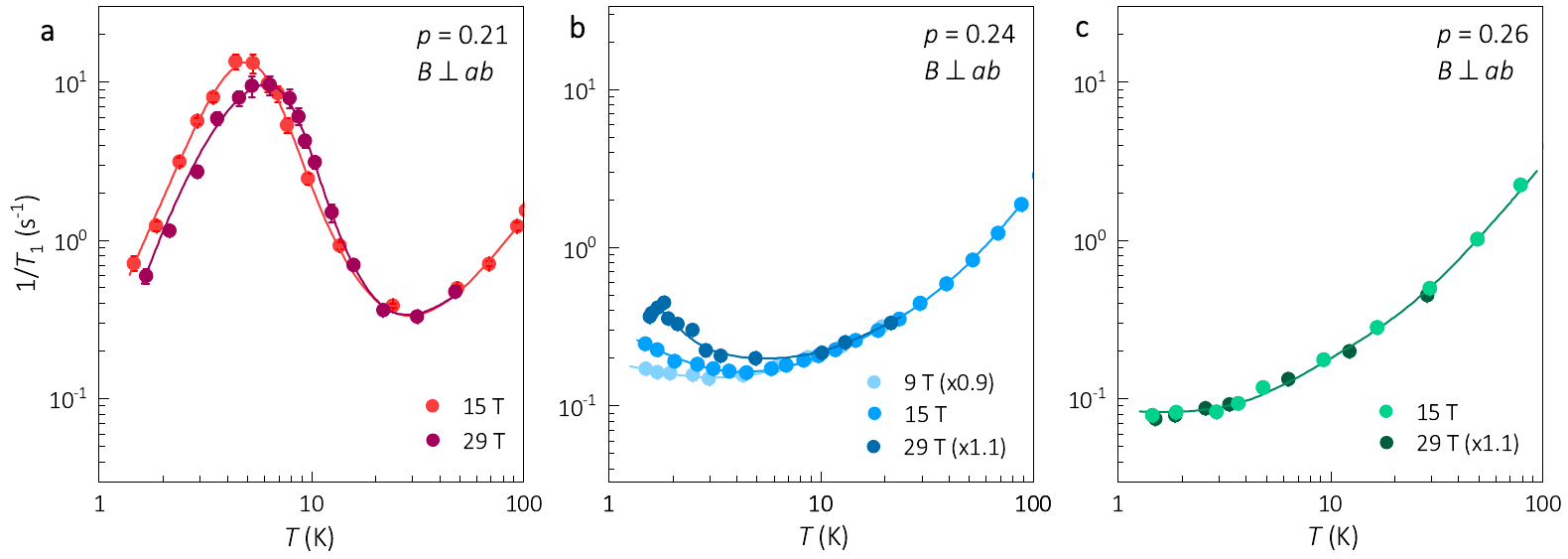}
  \caption{\label{T1field} Effect of field strength on spin dynamics. $1/T_1$ vs. temperature for $B\perp ab$ plane for $p=0.21$ (a), $p=0.24$ (b) and $p=0.26$ (c). The scaling factor between data at different field values presumably arises from different experimental conditions: the frequency window excited by the NMR pulse (which is smaller than the line width) is about field independent whereas the line width is field dependent, thus the fraction of excited nuclei varies with field. The lines are guides to the eye. }
\end{figure*} 

\subsection{Effect of field strength on spin order}
 
We now study the effect of field strength, with $B$ applied perpendicular to the planes (Fig.~\ref{T1field}).

\begin{figure*}[t!]
 \includegraphics[width=18cm]{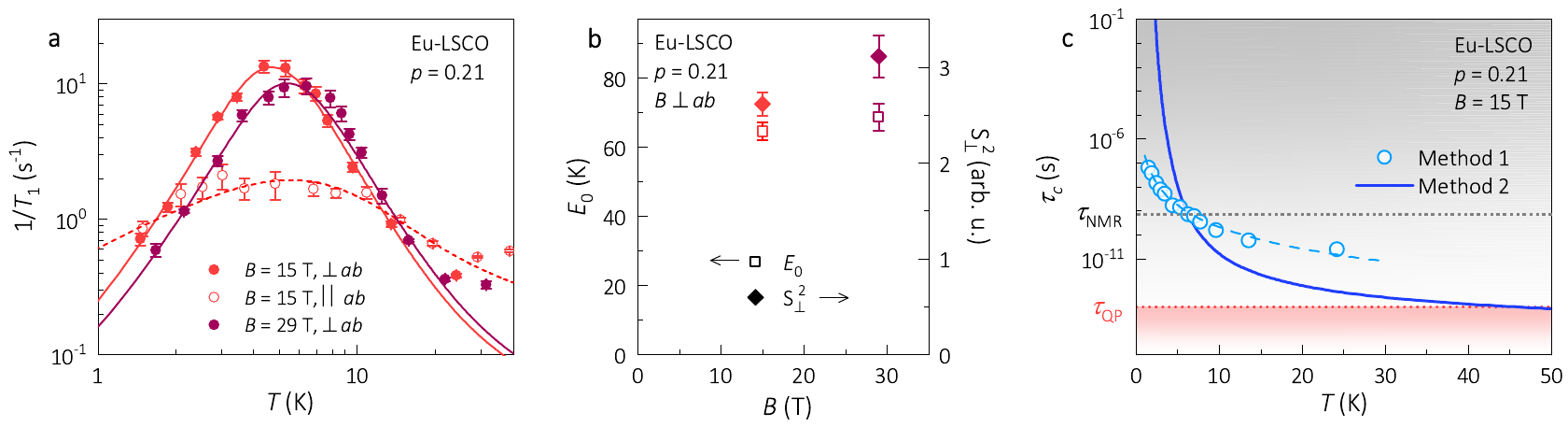}
 \caption{\label{fits} Quantifying the slow fluctuations. (a) Fit of $T_1$ data using Eq.~\ref{eq:BPP} and an exponential $T$ dependence of $\tau_c$ (Eq.~\ref{eq:tau}), with a distribution of parameters as in ref.~\cite{Vinograd2022}. Fitting of the broad flat-topped peak for $B\perp ab$ requires assuming very broad distributions of $E_0$ and $\tau_{\infty}$ values , resulting in $E_0$ being poorly constrained and the fitting model becoming questionable. (b) Fitting parameters $E_0$ and $S_\perp^2$ at 15 and 29~T for $B\perp ab$. For both fields, the distribution width $\Delta E_0/E_0\simeq 0.3$. (c) The correlation time $\tau_c$ of Eq.~\ref{eq:BPP}, extracted from the $T_1$ data at 15~T for $p=0.21$ in two different ways. Method~1 solves Eq.~\ref{eq:BPP} for $\tau_c$ (circles), ignoring any distribution of parameters. This yields approximately $\tau_c\propto T^{-3}$ (dashes). Method~2 corresponds to $\tau_c$ calculated using the $E_0$ value determined from the fit in (a).  $\tau_{\rm QP}$ is the quasiparticle lifetime, estimated to be of the order of 10$^{-13}$~s in ref.~\cite{Fang2022} and $\tau_{\rm NMR}$ defined by $\omega_{n} \tau_{\rm NMR}=1$ where $\omega_{n}$ is the NMR frequency.}
\end{figure*}

Below $p^*$ ($p=0.21$), the peak temperature $T_g^{\rm NMR}$ increases slightly from 15 to 29 T (Fig.~\ref{T1field}a). This is because we probe dynamics at an almost doubled NMR frequency compared to 15~T \textcolor{black}{and because $E_0$ (Eq.~\ref{eq:tau}) has increased (see below)}. On the other hand, the maximum amplitude of $1/T_1$ is  slightly reduced at 29 T (Fig.~\ref{T1field}a). This does not mean that the magnetic moments are reduced by the field: as discussed in section II, the value of $1/T_1$ at  $T_{\rm g}^{\rm NMR}$ is inversely proportional to the field, if the freezing itself is a field-independent processus. Therefore, the data indicate that $\altmathcal{S}_{\perp}(0)$ (see Eq.~\ref{eq:BPP}) has grown here by much less than a factor of 2 from 15 to 29~T. As a matter of fact, fitting the $1/T_1$ peak according to the procedure described in ref.~\cite{Vinograd2022} yields a mild increase for both $\altmathcal{S}_{\perp}(0)$ and $E_0$ \textcolor{black}{(see Fig.~\ref{fits})}. 

At or very near $p^*$ ($p=0.24$), the low $T$ upturn in $1/T_1$ is manifestly enhanced upon increasing field (Fig.~\ref{T1field}b), indicating that the field promotes fluctuations at the NMR frequency ({\it i.e.} slow spin fluctuations). Nevertheless, we do not know whether $1/T_1$ peaks at finite temperature or keeps increasing as $T\rightarrow0$. There may be a peak at approximately 1.8~K at 29~T but this impression could also result from scatter in the data points. However, whether a peak is present or not is ultimately unimportant: the essential point is that Eu-LSCO is still inclined to form spin stripes at ${\rm x}=0.24$, but in zero field, this is hindered as superconductivity sets in before quasi-static spin fluctuations can develop: the upturn in $1/T_1$ is seen only well below the zero-field $T_c(0)=12$~K at this doping (while it was visible above $T_c(0)=15$~K for $p=0.21$).

Above $p^*$ ($p=0.26$), the field strength has no effect whatsoever on 1/$T_1$ (Fig.~\ref{T1field}c). As previously mentioned, this insensitivity to the field may be expected due to the potential absence of bulk superconductivity at 15 T in both field orientations (Fig. ~\ref{tuning}f). However, earlier studies in LSCO have shown that slow spin dynamics remains field dependent even in conditions where superconductivity should be suppressed~\cite{Frachet2020,Vinograd2022,Frachet2021}. Therefore, if there were any weak underlying tendency towards freezing, one would expect a field dependence of 1/$T_1$ within the range of $B$ and $T$ explored here. Yet, this is not observed, indicating no sign of a tendency towards spin freezing above $p^*$. We cannot exclude that there remain small areas experiencing spin freezing in the sample but these must be insignificant in the overall behavior.

\section{Discussion}

\subsection{Relation to earlier magnetic measurements}

In the light of these results and of four consistent reports from NMR or $\mu$SR in Eu-LSCO~\cite{Klauss2000,Suh2000,Hunt2001,Grafe2010}, the absence of static magnetic order for $p=0.20$ doping in Nd-LSCO~\cite{Nachumi1998} appears to be singular. It is in fact likely that the wide distribution of ordered moments and of freezing temperatures  in zero field have made the $\mu$SR detection challenging in the latter work. In addition, any slight oxygen over-stoichiometry could have further broadened the distribution, decreased the freezing temperature and possibly lowered the magnetic volume fraction, all factors being susceptible to reduce the magnetic signal below the detection threshold. 

We also note that the absence of spin freezing in our $B \parallel ab$ data for $p=0.24$ contradicts the neutron scattering report of spin-stripe order in zero field at this doping in Nd-LSCO~\cite{Ma2021}. The standard explanation for such a discrepancy is that the order is static at the neutron timescale but not at the NMR timescale. However, this explanation seems unlikely here, as our $B \parallel ab$ data show no evidence of spin fluctuations slowing down. Instead, the ordering and/or slow fluctuations of the Nd moments may have contaminated the elastic response in the neutron experiment, \textcolor{black}{or this elastic response could originate from a fraction of the sample too small to be detected in our $T_1$ measurements}.

\subsection{The singular case of $p=0.24$}

\begin{figure*}[t!]
 \includegraphics[width=17cm]{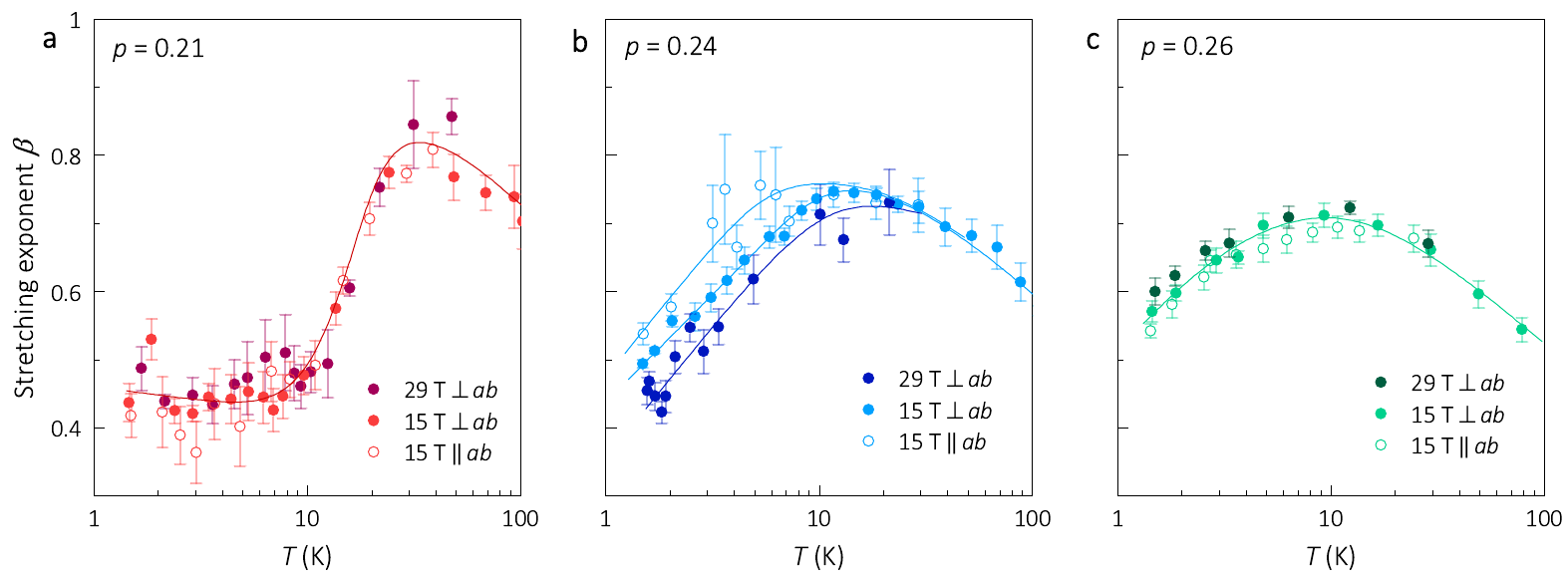}
  \caption{\label{stretching} Temperature dependence of the stretching coefficient $\beta$. $\beta$ deviates from 1 both at low temperature because of a spatial distribution of $T_1$ values and at high temperature because of quadrupolar relaxation (temporal fluctuations of the electric field gradient at structural transitions, see Appendix~\ref{Structure} for further details). The lines are guides to the eye. \textcolor{black}{For $p=0.21$ (a), $\beta$, which is a non-linear function of the width of the distribution of $T_1$ values~\cite{Johnston2006}, is slightly smaller for $B\parallel ab$ than for $B\perp ab$ at low $T$. This suggests stronger spatial inhomogeneity due to stronger superconductivity in this $B \parallel ab$ configuration (see also related discussion in Fig.~\ref{fits}).}}
\end{figure*}

At $p=0.24$ doping, we observe remnant slow, inhomogeneous spin-stripe fluctuations that emerge only after superconductivity is weakened by a perpendicular field (Figs.\ref{T1angle}c, \ref{T1field}c). However, this effect is considerably weaker than for $p=0.21$ (see direct comparison of $1/T_1$ data in Fig.~\ref{comp}, where we also show data in LSCO at different doping levels for comparison), and most of the spins possibly continue to fluctuate slowly down to $T=0$. Increasing the doping from $p=0.21$ to $p=0.24$ significantly reduces the amount of slow fluctuations on average, but the level of spatial inhomogeneity at base temperature remains mostly unchanged (Fig.\ref{stretching}a,b). This inhomogeneity actually remains substantial at $p=0.26$ (Fig.~\ref{stretching}c), even if there is no sign of freezing in $1/T_1$ vs. $T$ at this doping level.

The sluggish dynamics for $p=0.24$ is consistent with this doping being just on the verge of freezing, that is, static spin-stripe order ending at $p\simeq p^*$. The persistent inhomogeneity across $p^*$, on the other hand, shows that freezing and inhomogeneity are not necessarily tied to each other. Nevertheless, gaining further insights into real space would require unequivocal information on the $T_1$ distribution, which represents an important challenge for future work. An appealing, though perhaps not unique, interpretation of our data relies on nanoscale phase separation between stripe-ordered puddles and homogeneous regions without order~\cite{Li2022,Tranquada2024}. In this scenario, the detected slow fluctuations originate predominantly from the striped puddles and it is possible that freezing occurs as long as these puddles percolate or remain sufficiently coupled. Beyond a certain doping level, proposed to be $p^*$ in refs.~\cite{Li2022,Tranquada2024}, the striped puddles may become too small and/or too sparse to sustain widespread freezing but electronic inhomogeneity remains over some range of doping.

\subsection{Intimate connection between spin-stripe order and the pseudogap phase}
\label{connection}

\begin{figure}[t!]
 \includegraphics[width=8cm]{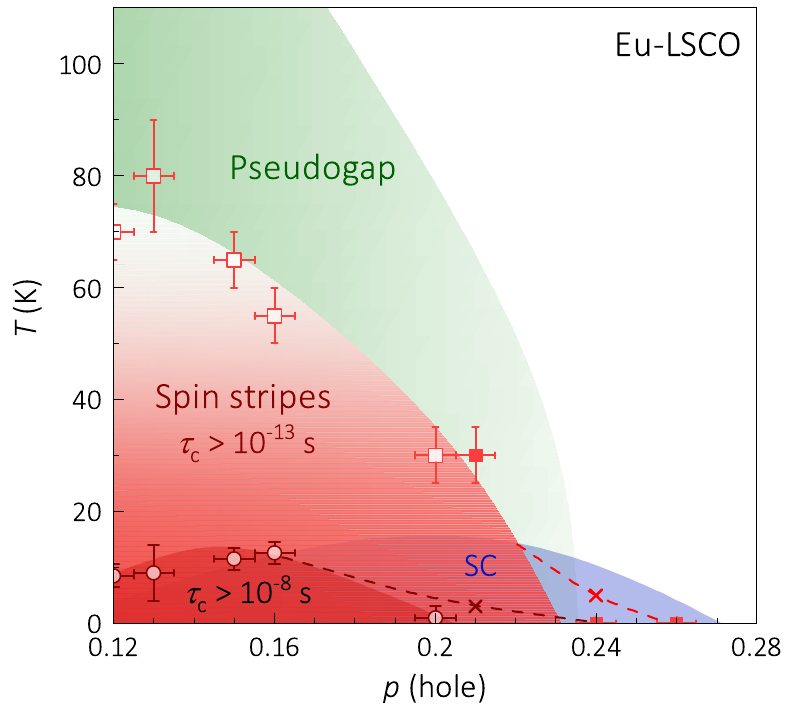}
  \caption{\label{spinstripes} Magnetic phase diagram in zero field. Circles correspond to the temperature of freezing at the NMR or $\mu$SR timescale. Squares correspond to the temperature at which spin fluctuations become slower than 10$^{-13}$~s, as approximately estimated from the temperature at which $1/T_1$ starts its low $T$ upturn. This typical temperature is where transport properties are susceptible to be impacted by the slow spin fluctuations. Data for $p\leq0.20$ (open symbols) are from refs.~\cite{Suh2000,Hunt2001,Curro2000}. Filled symbols are zero-field extrapolations based on the present study, while the crosses correspond to our high-field results. The dashed lines indicate how the boundaries approximately shift upon applying a large magnetic field. The pseudogap boundary is according to~\cite{Cyr2018} and references therein. }
\end{figure}

We come to the main point of the paper. Strictly speaking, the data show glassy spin freezing ending in-between $p=0.21$ and 0.24 in zero field and very close to $p=0.24$ in high fields (see phase diagram, Fig.~\ref{spinstripes}). In both cases, this is very near $p^*$. Having argued that the distribution of local doping  in real compounds calls for caution when discussing precise doping values in the phase diagram of La214 cuprates, we take these results as evidence that spin-stripe order, and thus the intertwinement of spin and charge stripes, is intrinsically ({\it i.e.} in the hypothetical absence of inhomogeneity) present up to $p^*$, and not above, in non-superconducting Eu-LSCO. 

This conclusion parallels earlier results in LSCO where spin-stripe order was also found to end at $p^*$ in the absence of superconductivity~\cite{Frachet2020,Vinograd2022}. The difference between LSCO and Eu-LSCO lies in the strength of superconductivity, which determines the fierceness of the competition with stripe order and, in turn, results in different values of the maximum doping ($p_c$) at which the ground state is spin ordered in zero-field. In LSCO, superconductivity pushes back $p_c$ from $p^*=0.19$ (the value in high fields) down to $p=0.135$ (the value in zero field)~\cite{Frachet2020}, while in Eu-LSCO, $p_c$ shifts only slightly, if it shifts at all. 

While it is always possible to argue that the end of spin-stripe order at $p^*$ is a mere coincidence in a given compound, this critique is no longer tenable now that the same conclusion has been reached in two systems with notably different balances between superconductivity and stripe order. The overarching conclusion is therefore that the pseudogap and stripe phases are closely linked in La214 cuprates. This conclusion aligns with recent theoretical work by \v{S}imkovic {\it et al.}~\cite{Simkovic2024} who showed that, regardless of the value of $U/t$ in the one-band Hubbard model, the pseudogap phase consistently terminates at the same doping level as the stripe phase~\cite{Xu2022}. We note, however, that the relationship between the two phenomena may not be causal. As discussed in Refs.~\cite{Frachet2020,Vinograd2022}, both the stripes and the pseudogap may arise as inevitable consequences of doping a 3$d^9$ Mott insulator on the square lattice, \textcolor{black}{both ultimately vanishing upon increasing doping as Mott physics becomes irrelevant}.

\subsection{Fermi-surface reconstruction below $p^*$}

\textcolor{black}{In principle, the quantum phase transition from a striped to an uniform magnetic ground state could account for the enhanced specific heat coefficient seen around $p^*$ in Eu/Nd-LSCO~\cite{Michon2019}. The spin order that we see at $p=0.21$ is also susceptible to reconstruct the Fermi surface~\cite{Millis2007}, thus potentially explaining the ADMR (angle-dependent magnetoresistance) and Hall effect results~\cite{Fang2022,Collignon2017}. The main difficulties with such scenario have been spelled out by Taillefer and coworkers (without ruling it out, though)~\cite{Michon2019,Collignon2017,Fang2022}: how could reconstruction arise from an order that is neither truly long-ranged nor fully static and how could this scenario explain that YBa$_2$Cu$_3$O$_{\rm y}$ has a similar doping dependence of the Hall constant as Nd/Eu-LSCO~\cite{Badoux2016a} but no spin-stripe order below $p^*$~\cite{Wu2011}? This question is important because, if spin-stripe order does not significantly transform the Fermi surface, the low-temperature properties of Eu/Nd-LSCO can be considered representative of the generic pseudogap Fermi surface -- {\it i.e.}, the Fermi surface of the pseudogap state without the spin-stripe order specific to La214. This would, in turn, lend a universal character to the conclusions drawn from studies on this singular class of cuprates.}

\textcolor{black}{As discussed above from a qualitative perspective, since the fluctuations at low temperatures reach the $\mu$eV range (and probably even lower), the spin-stripe order can effectively be considered static at $p=0.21$ for all practical purposes. We now turn to a more quantitative analysis of the fluctuation timescale, as a function of temperature.}

It has been proposed that electrical transport is affected by spin order as soon as the staggered moments appear static on the timescale of a quasiparticle lifetime $\tau_{\rm QP}$ of 10$^{-13}$~s~\cite{Fang2022,Vinograd2022}, {\it i.e.} as soon as fluctuations become slower than 10~THz $\sim$~40~meV. The question is then how long $\tau_c$ is in the conditions of the transport experiments, namely  at 33~T and $T\rightarrow 0$ for the Hall effect~\cite{Collignon2017} and 45~T and 25~K for the ADMR experiment~\cite{Fang2022}.

We determined $\tau_c$ for $p=0.21$ in two different ways, using the $B\perp ab = 15$~T data (the 29 T data provide nearly identical results). First, we simply solved Eq.~\ref{eq:BPP} for $\tau_c$, which has the advantage of not assuming any $T$ dependence but has the drawback of neglecting any distribution of $\tau_c$ values. As Fig.~\ref{fits}c shows, the $\tau_c$ values thus determined are longer than $\tau_{\rm QP} =10^{-13}$~s for $T \lesssim 30~K$ (the analysis does not apply at higher $T$ for there is no sign of slowing down). Second, we fitted the peak in $1/T_1$  using the same method as in ref.~\cite{Vinograd2022}, namely assuming an exponential growth of $\tau_c$ and a distribution of each parameter in Eq.~\ref{eq:tau}. Again, we find $\tau_c$ values (Fig.~\ref{fits}c) that are larger than $\tau_{\rm QP}=10^{-13}$~s, already at $T\simeq30$~K. Therefore, we conclude that, for $p=0.21$ in Eu/Nd-LSCO, the transport experiments have been performed in conditions for which spin degrees of freedom may be considered as static. For $p=0.24$, our observation of field-induced slow fluctuations does not seem to align with the unreconstructed Fermi surface deduced from high-field ADMR results in Nd-LSCO ~\cite{Fang2022}. However, the field-induced magnetism appears at much lower temperature ($\sim 6$~K, Fig.~\ref{T1field}b) than the temperature (25~K) at which ADMR was measured so there is no actual discrepancy. 

What about the correlation length? As a local technique, NMR does not directly estimate the coherence length. According to neutron scattering, the in-plane correlation length $\xi$ of quasi-elastic scattering from spin-stripe order is around $100$~\AA~for Nd-LSCO $p\simeq 0.20$ in zero field~\cite{Ma2021}. \textcolor{black}{This is already greater than the CDW correlation length in underdoped cuprates, except for long-range CDW order near $p=0.12$, in either YBCO in high fields or Eu/Nd-LSCO La$_{2-{\rm x}}$Ba$_{\rm x}$CuO$_4$}. Furthermore, $\xi$ should increase with field: in La$_{2-{\rm x}}$Sr$_{\rm x}$CuO$_4$ (${\rm x}=0.144$) which shows no order in zero field, a  3~T field is sufficient to induce magnetic Bragg peaks with $\xi>120$~\AA~\cite{Khaykovich2005}. \textcolor{black}{Therefore, the correlation length of spin-stripe order should not be a limiting factor for Fermi surface reconstruction in Nd/Eu-LSCO. Neutron scattering measurements in sufficiently high fields would be valuable to confirm this conjecture.}

\textcolor{black}{From the discussion above, we conclude that the Fermi surface in the stripe-ordered state is unlikely to be identical to the "generic pseudogap Fermi surface". In other words, low-$T$ experiments in La214 impose constraints on the pseudogap Fermi surface but do not probe it directly. While determining the precise Fermi surface of the pseudogap state is beyond the scope of this paper, we emphasize that stating that the Fermi surface is reconstructed into small closed pockets in the stripe-ordered state does not imply that the Fermi surface is entirely different (that is, open and large) in the absence of stripe order. It may consist of pockets as well, though not necessarily of the same size.}

\subsection{\textcolor{black}{Connection to} resistivity upturns}

It has been previously noted that resistivity upturns coincide with the emergence of quasi-static spin fluctuations (ref.~\cite{Vinograd2022} and references therein). Consistent with this phenomenology, the in-plane resistivity $\rho_{ab}$ at 33~T for $p=0.21$, in both Eu-LSCO~\cite{Michon2019} and Nd-LSCO~\cite{Collignon2017}, shows an upturn below 30~K, precisely where the slow spin-fluctuations become evident in the $T_1$ data. This correlation thus supports the idea that quasi-static spin fluctuations have a significant impact on electrical transport (see also~\cite{Campbell2024}).

\textcolor{black}{Conversely, at $p=0.24$, while $1/T_1$ upturns below $\sim 6$~K in high fields, no concomitant resistivity upturn is observed in our measurements~\cite{LeBoeuf2025} as well as in previous studies of both Eu-LSCO~\cite{Michon2019} and Nd-LSCO~\cite{Michon2019,Collignon2017}. This might be seen as contradicting the above described phenomenology. However, it is likely that the spectral weight of quasi-static spin fluctuations at this doping is too weak to affect $\rho_{ab}$ in a measurable way, at least in the range of fields and temperatures of the transport measurements: $T\geq 5$~K (2~K) in Eu-LSCO (Nd-LSCO) and $B= 33$~T). It is possible that an upturn would be seen in measurements performed at higher fields and/or lower temperatures. }

\subsection{\textcolor{black}{Connection to} strange metal behavior}

The $p=0.24$ doping is particularly intriguing as, in the absence of an upturn, the resistivity is linear from 80~K down to at least 2~K (ref.~\cite{Collignon2017} \textcolor{black}{and unpublished measurements in our own sample~\cite{LeBoeuf2025}}). This so-called strange metal behavior~\cite{Phillips2022} seems to emerge in the regime where the system may be viewed as a kind of 'incipient glass' with spatially-heterogeneous spin fluctuations being present at low, but not quasi-static, frequencies (after superconductivity has been suppressed). Given that gapless fluctuations and spatial disorder are main ingredients of strange metallicity in several theoretical works~\cite{Cha2020,Wu2022,Ciuchi2023,Caprara2022,Patel2023,Patel2024,Fratini2024}, we hypothesize that the unusual magnetic behavior observed here at $p=0.24$ may underlie the strange metal behavior. \textcolor{black}{A similar conclusion has been reached recently from high-field transport measurements in LSCO~\cite{Campbell2024}.}

\textcolor{black}{Linear-in-$T$ resistivity is also observed for our $p=0.26$ sample~\cite{LeBoeuf2025} where} the saturation of $1/T_1$ as $T \rightarrow 0$ could be interpreted as a remnant of these heterogeneous sluggish fluctuations.

\textcolor{black}{\section{Conclusion and outlook}}

\textcolor{black}{By studying low-energy spin dynamics in Eu-LSCO as a function of temperature, magnetic field and doping, we found that spin-stripe order exists, and competes with superconductivity, up to a doping level consistent with the pseudogap critical doping $p^*$. In contrast to LSCO~\cite{Frachet2020,Vinograd2022}, suppressing superconductivity in high fields does not significantly expand the doping range over which spin-stripe order exists. Above $p^*$, dynamical spin-stripe correlations persist but do not freeze at low temperatures. These findings suggest a fundamental, yet to be fully elucidated, connection between the pseudogap phase and the stripe phase with intertwined spin and charge orders.}

\textcolor{black}{In this NMR study, we were unable to draw conclusions about the presence or absence of charge-stripe order, for reasons detailed in Appendix~\ref{AboutCDW}, including the difficulty of defining a meaningful CDW onset temperature. Whether static charge-stripe order terminates below, at, or above $p^*$ remains an important open question in La214, as well as in other cuprate families (see refs.~\cite{Fujita2014,Lu2022,Peng2018,Li2021,Song2023} for varying observations in Bi-based cuprates). Experiments in high magnetic fields will certainly be useful for addressing this question. Nonetheless, our findings already imply that if a static CDW persists above $p^*$ in La214, as refs.~\cite{Miao2021,Li2023} suggest, it must be decoupled from the spin order. An alternative possibility is that spin and charge correlations remain intertwined above $p^*$ but  become both dynamical.}

\textcolor{black}{In our view, a crucial question is whether disorder and electronic inhomogeneity play a central role in the observed phenomenology. This encompasses the proposal that striped/pseudogapped and uniform regions phase separate over some length scale, and that many of the changes observed at $p^*$ can be understood in terms of a percolation transition, at least in La214~\cite{Li2022,Tranquada2024}. We hope that our results will stimulate predictions for NMR and other experimental probes within theoretical models that fully address these questions.}

\vspace{1cm}
\begin{acknowledgments}

\textcolor{black}{We are grateful to L. Chaix, J. Chang, N.B. Christensen, M. Frachet, M. H\"{u}cker, S. Fratini, B. Gaulin, I.~Vinograd for valuable exchanges.}

This work was performed at the LNCMI, a member of the European Magnetic Field Laboratory. Work at LNCMI was supported by the Laboratoire d'Excellence LANEF (ANR-10-LABX-51-01) and by the French Agence Nationale de la Recherche (ANR) under reference ANR-19-CE30-0019 (Neptun). 

\end{acknowledgments}

\appendix

\section{Samples}

The single crystals were grown by the travelling solvent floating zone method in Tokyo ($p=0.21$ and $0.24$) and in Hefei ($p=0.26$). Crystals from the same Tokyo batches have been previously studied by other probes (see~\cite{Michon2019,Laliberte2015,Cyr2018,Chen2024,Kuspert2022,vonArx2023} and references therein). To the best of our knowledge, $p=0.26$ has not been studied before in Eu-LSCO. 

   \begin{figure*}[b!]
 \includegraphics[width=15cm]{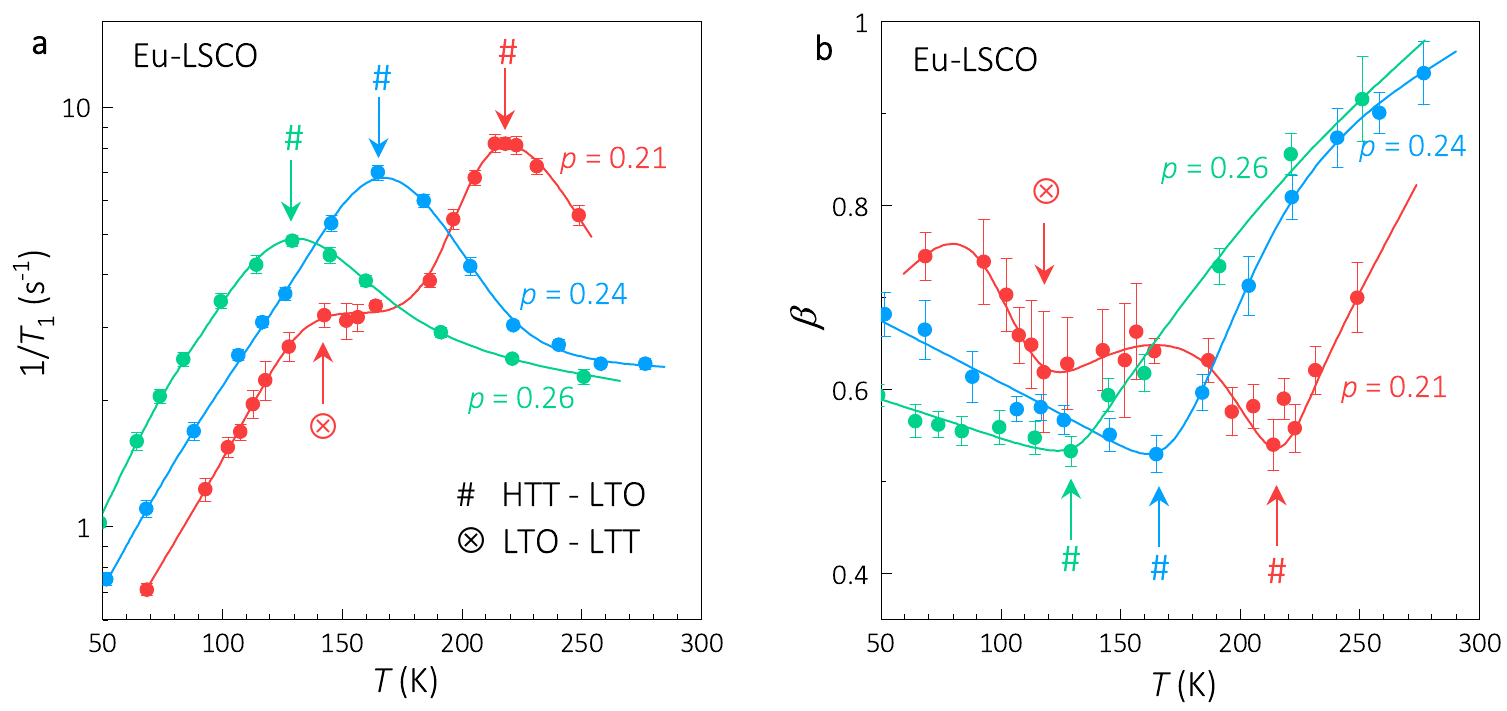}
  \caption{\label{htt} Structural phase transitions HTT to LTO and LTO to LTT, measured as anomalies in (a) the $T$ dependence in $1/T_1$ and (b)  the stretching exponent $\beta$ in Eq.~\ref{recovery}. The corresponding temperatures, reported in the phase diagram of Fig.~\ref{phasediag}, are consistent with literature data.}
\end{figure*} 

\section{NMR methods}
\label{Methods}
We used home-built NMR probes and spectrometers. Experiments up to 15~T were performed in a superconducting magnet and experiments at 29~T were performed in the LNCMI M10 resistive magnet. $^{139}$La NMR spectra were obtained by the frequency-shifted and summed Fourier transform technique of Clark {\it et al.}~\cite{Clark1995}. The relaxation rate $T_1^{-1}$ was measured on the central transition of $^{139}$La. 

We have taken particular care to use low radio-frequency power to avoid transient heating at low temperatures. This issue is common in cuprate single crystals, and indeed already identified in our previous LSCO works~\cite{Frachet2020,Frachet2021,Vinograd2022}.

\section{Data analysis}
\label{Analysis}
The $T_1$ values were determined by fitting the recoveries (time $t$ dependence of the nuclear magnetization  $M(t)$) to a stretched version of the theoretical law for magnetic relaxation between $m_I=\pm 1/2$ levels of a nuclear spin $I=7/2$~\citep{Mitrovic2008,Arsenault2020,Frachet2020}:
\begin{multline}
M(t) =  M_0 \Big[ 1-a\Big(0.714 \, e^{-\left({28t \over T_1}\right)^\beta} - \, 0.206\, e^{-\left({15t \over
T_1}\right)^\beta}\\
 - \, 0.068 \, e^{-\left({6t \over T_1}\right)^\beta}  - \,0.012\, e^{-\left({t \over T_1}\right)^\beta} \Big)\Big] ,
\label{recovery}
\end{multline}
where the coefficients $a$ (ideally $a=2$ for a perfect inversion, which could not be reached here due to the narrow excitation width) and $M_0$ (equilibrium nuclear magnetization) were also fit parameters.

\section{Structural transition temperatures}
\label{Structure}
The structural transition temperatures were determined by $^{139}$La measurements (Fig.~\ref{htt}): temporal fluctuations of the electric field gradients produce a peak in the $T$ dependence of $1/T_1$ (through quadrupole relaxation)~\cite{Suh2000,Baek2013,Frachet2020} and because a deviation from one of the stretching exponent $\beta$ as we fit with Eq.~\ref{recovery} which is not appropriate for quadrupolar relaxation.

While all three samples show a transition from the high temperature tetragonal (HTT) phase to the low temperature orthorhombic (LTO), only the $p=0.21$ sample shows an additional LTT phase at low temperature. 

\section{What about charge stripes?}
\label{AboutCDW}

Unlike in YBa$_{2}$Cu$_{3}$O$_y$~\cite{Wu2011,Wu2015,Vinograd2021} and Bi2201~\cite{Kawasaki2017}, CDW order is not detected directly by NMR in La214 cuprates: $^{17}$O and $^{63}$Cu nuclei experience such fast relaxation by spin-stripe fluctuations that the NMR-signal intensity is wiped out~\cite{Hunt1999,Curro2000,Julien2001} and $^{139}$La nuclei, which are less strongly coupled to the CuO$_2$ planes, are apparently not sensitive enough for unambiguously detecting charge order. Nonetheless, indirect NMR evidence of CDW order in La214 has previously been discussed in relation with the following observation: slow fluctuations become visible in NMR at a temperature that matches the temperature at which superlattice peaks appear in X-ray diffraction. The notion that charge ordering triggers the slow spin fluctuations has then led to use the onset of slow spin fluctuations as a proxy for the CDW onset temperature $T_{\rm CDW}$ (\cite{Vinograd2022} and refs. therein).
   \begin{figure*}[t!]
 \includegraphics[width=15cm]{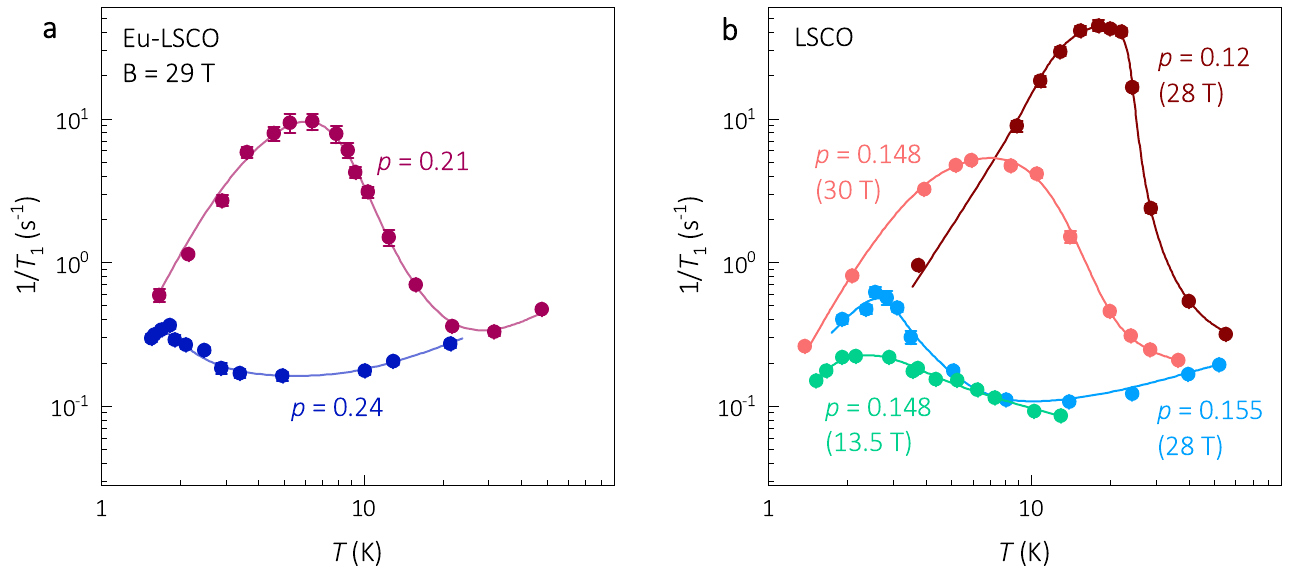}
  \caption{\label{comp} Direct comparison of spin-stripe freezing from $1/T_1$ measurements in La214 compounds. (a) Eu-LSCO $p=0.21$ and $0.24$ at the same field of 29~T (data from Fig.~\ref{T1field}). (b) LSCO at various doping and field values (data from refs.~\cite{Frachet2020,Frachet2021,Vinograd2022}). }
\end{figure*} 
 \begin{figure}[b!]
 \includegraphics[width=8cm]{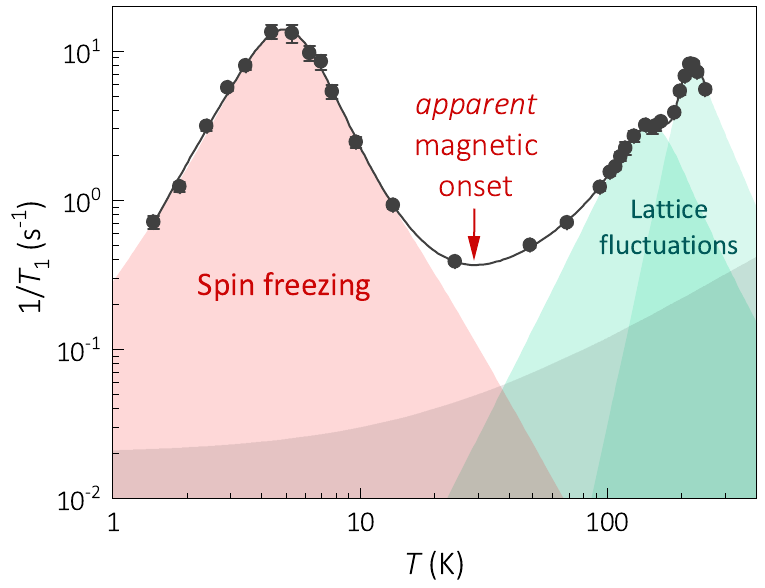}
  \caption{\label{onset} Sketch of contributions to $1/T_1$ for Eu-LSCO $p=0.21$ illustrating that the temperature of 30~K at which $1/T_1$ has a minimum is only an apparent onset temperature of magnetic freezing: the real onset is possibly masked by quadrupolar relaxation, due to structural transitions at higher temperatures. The grey shading represents a, supposedly metallic-like, background relaxation $a+b\, T$.}
\end{figure} 

This picture must, however, now be partly revised. Indeed, the newest X-ray scattering experiments have revealed that the CDW transition temperatures from previous experiments were actually biased by the experimental sensitivity~\cite{Wang2020,Lee2022}. The new data with improved sensitivity do not show a well-defined onset temperature. In fact, the question of whether a physically meaningful onset temperature can be defined does not only concern La214 but is entirely general: it is true of any CDW in the presence of quenched disorder~\cite{Nie2014,Wu2015,Straquadine2019}, which includes the short-range CDW phase of the cleaner cuprate YBCO~\cite{Wu2015,Vinograd2019}.

If no sharp onset for CDW formation can be defined, does the slowing down of spin fluctuations also lack a sharp onset? Or does the spin-freezing process still have a relatively clear onset when the CDW reaches a certain strength? Unfortunately, these questions cannot be definitively answered here, as the NMR data does not provide a perfectly unambiguous definition of the onset, as illustrated in Fig.~\ref{onset}. The temperature at which $1/T_1$ begins its low $T$ upturn serves as a reasonable indication of when quasi-static spin fluctuations become prominent. Even if this temperature does not represent a well-defined onset, that it extrapolates to zero near $p^*$ (Fig.~\ref{spinstripes}) is consistent with the end of spin-stripe order at $p^*$ and the same doping dependence would presumably be observed if another criterion had be chosen. However, based on the above discussion, we refrain from concluding that static CDW order terminates at $p^*$.

\bibliography{EuLSCObib}

\end{document}